\documentclass[transmag]{IEEEtran}
\usepackage{latexsym}
\usepackage{graphicx}
\usepackage{amsfonts,amssymb,amsmath}
\usepackage{hyperref}
\usepackage{epsfig}
\usepackage{setspace}
\usepackage{booktabs}
\usepackage{tabularx}
\usepackage{epstopdf}
\usepackage{url}
\usepackage{xcolor}
\usepackage{balance}
\newcommand{\etal}{{\em et al. }}       
\newcommand{\etalns}{{\em et al.}}
\newcommand{\eg}{{\em e.g.}}           
\newcommand{\ie}{{\em i.e.}}           
\newcommand{\re}[1]{\textcolor{black}{#1}}
\newcommand{\rev}[1]{\textcolor{black}{#1}}

\def\BibTeX{{\rm B\kern-.05em{\sc i\kern-.025em b}\kern-.08em T\kern-.1667em\lower.7ex\hbox{E}\kern-.125emX}}

\begin{document}

\title{A novel multiple instance learning framework for COVID-19 severity assessment via data augmentation and self-supervised learning}

\author{Zekun Li$^\dagger$, Wei Zhao$^\dagger$, Feng Shi, Lei Qi, Xingzhi Xie, Ying Wei, Zhongxiang Ding,\\ Yang Gao, Shangjie Wu, Jun Liu$^\star$, Yinghuan Shi$^\star$, Dinggang Shen$^\star$
\thanks{$\dagger$ Zekun Li and Wei Zhao are the co-first authors.} 
\thanks{* Asterisk denotes co-corresponding authors.}
\thanks{Zekun Li, Yinghuan Shi, and Yang Gao are with the State Key Laboratory for Novel Software Technology, Nanjing University, China. They are also with National Institute of Healthcare Data Science, Nanjing University, Nanjing, 210046, China.}
\thanks{Lei Qi is with School of Computer Science and Artificial Intelligence, Southeast University, Nanjing, 210018, China.}
\thanks{Ying Wei, Feng Shi, and Dinggang Shen are with Department of Research and Development, Shanghai United Imaging Intelligence Co., Ltd., Shanghai, 201807, China. Dinggang Shen is also with School of Biomedical Engineering, ShanghaiTech University, Shanghai, China and Department of Artificial Intelligence, Korea University, Seoul 02841, Republic of Korea.}
\thanks{Wei Zhao, Xingzhi Xie, and Jun Liu are with Department of Radiology, The Second Xiangya Hospital, Central South University, Hunan, 410011, China. Jun Liu is also with Department of Radiology Quality Control Center, Changsha, 410011, China.}
\thanks{Zhongxiang Ding is with the Department of Radiology, Hangzhou First People’s Hospital, Zhejiang University School of Medicine, Hangzhou, Zhejiang, China.}
\thanks{Shangjie Wu is with Department of Pulmonary and Critical Care Medicine, The Second Xiangya Hospital, Central South University, Changsha, 410011, China.}}

\IEEEtitleabstractindextext{\begin{abstract}How to fast and accurately assess the severity level of COVID-19 is an essential problem, when millions of people are suffering from the pandemic around the world. Currently, the chest CT is regarded as a popular and informative imaging tool for COVID-19 diagnosis. However, we observe that there are two issues -- weak annotation and insufficient data that may obstruct automatic COVID-19 severity assessment with CT images. To address these challenges, we propose a novel three-component method, \ie, 1) a deep multiple instance learning component with instance-level attention to jointly classify the bag and also weigh the instances, 2) a bag-level data augmentation component to generate virtual bags by reorganizing high confidential instances, and 3) a self-supervised pretext component to aid the learning process. We have systematically evaluated our method on the CT images of 229 COVID-19 cases, including 50 severe and 179 non-severe cases. Our method could obtain an average accuracy of 95.8\%, with 93.6\% sensitivity and 96.4\% specificity, which outperformed previous works.\end{abstract}


}

\maketitle

\section{INTRODUCTION}

\label{sec1}
Recently, a new coronavirus, named by the World Health Organization (WHO) as COVID-19, has been rapidly spreading worldwide. As of 23 October 2020, there have been more than forty million confirmed COVID-19 cases globally. In view of its emergency and severity, WHO has announced COVID-19 outbreak a pandemic.

\begin{figure}[htbp]
\centerline{\includegraphics[scale=.45]{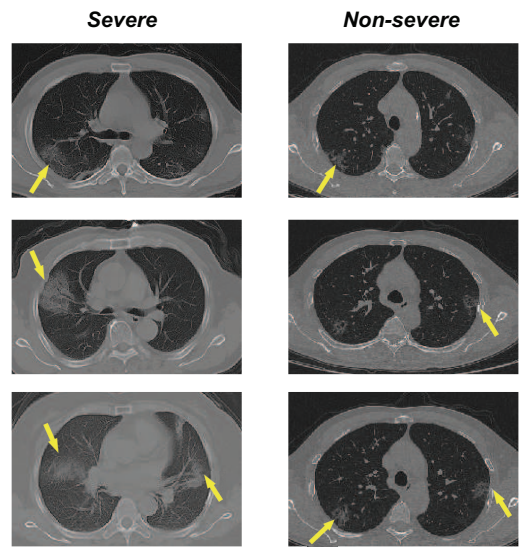}}
\caption{Examples of chest CT images with severe infection (left) and non-severe infection (right) of COVID-19. The yellow arrows indicate representative infection regions.}
\label{fig0}
\end{figure}

Due to the rapid spread, long incubation period and severe respiratory symptoms of COVID-19, clinical systems around the world are under tremendous pressure in multiple aspects. In current study, how to fast and accurately diagnose COVID-19 and assess its severity has become an important prerequisite for clinical treatment.

At present, for the diagnosis of COVID-19, traditional reverse transcription polymerase chain reaction (RT-PCR) is widely employed worldwide as a gold standard. However, due to its high false negative rate, repeat testing might be needed to achieve an accurate diagnosis of COVID-19. The chest computed tomography (CT) has been an imaging tool frequently used for diagnosing other diseases, and because it is fast and easy to operate, it has become a widely used diagnosis tool for COVID-19 in China. However, not only is manual diagnosis by CT images laborious, it is also prone to the influence of some subjective factors, \eg, fatigue and carelessness.

\rev{CT images are rather informative, with numbers of discriminative imaging biomarkers, so they are useful in assessing the severity of COVID-19. Based on the observation of Tang \etal  \cite{ra78}, the CT images of severe cases usually have larger volume of consolidation regions and ground glass opacity regions, than those of non-severe cases.} Therefore, several computer-aided methods \cite{ra78,yang2020chest,shan2020abnormal} have been recently proposed. However, we notice that current studies have neglected two important issues.
\begin{itemize}
\item \textbf{Weak Annotation}. Usually, it is rather time-consuming for physicians to precisely delineate the infection regions manually. Therefore, only the image-level annotation (\ie, label) for indicating the class of cases (\ie, severe or non-severe) is available, which could be regarded as a weakly-supervised learning setting. \textbf{This inspires us to develop a model that works under weakly-supervised setting (merely with image-level annotation)}.
\item \textbf{Insufficient Data}. According to current studies, it remains difficult to collect and label a large set of COVID-19 data. Besides, given the prevalence rate, the number of non-severe cases is much larger than that of severe cases, bringing about a significant issue of class imbalance, that raises the challenge of learning a stable model by avoiding overfitting. \textbf{This motivates us to seek ways to ease the imbalance of different classes and make the most of the insufficient data.}
\end{itemize}

Therefore, aiming to achieve the fast and accurate COVID-19 severity assessment with CT images, we propose a novel weakly supervised learning method via multiple-instance data augmentation and self-supervised learning. Our method is designed to solve the problems of weak annotation and insufficient data in a unified framework. 

On one side, the concept of \emph{weak annotation} comes from weakly supervised learning paradigm, whose goal is to develop learning models under three types of supervision: inexact supervision, incomplete supervision, and inaccurate supervision. In the severity assessment task, weak annotation is one type of inexact supervision where only the image-level label is provided by physicians whereas the region-level label is unavailable. Formally, we model this weak annotation task under multiple instance setting: We divide a CT image into several patches (\ie, unannotated instances), to make it as a bag consisting of multiple instances. Similar to multiple instance learning setting, the image indicated with severe or non-severe infection is considered as the positive or negative bag, respectively. 

On the other side, the problem of \emph{insufficient data} greatly challenges the robustness and stability of a learned model. We notice that in current studies on COVID-19, sizes of samples are often small. To tackle this challenge, we are motivated by two major aspects: 1)  to complement the original data by generating additional ``virtual'' data by using data augmentation technique, and 2) to leverage the patch-level information to benefit the learning process, since the quantity of patches is much larger than that of images. In particular, 1) we develop a simple yet effective multiple-instance data augmentation method to generate virtual bags to enrich the original data and guide stable training process; 2) Along with the bag-level labels for supervised training, there is also abundant unsupervised information that we can mine from the sufficient unannotated instances, so we apply self-supervised learning, in the form of patch location tasks, to exploit characteristic information of the patches. 

\begin{figure*}[htbp]
\centerline{\includegraphics[scale=.45]{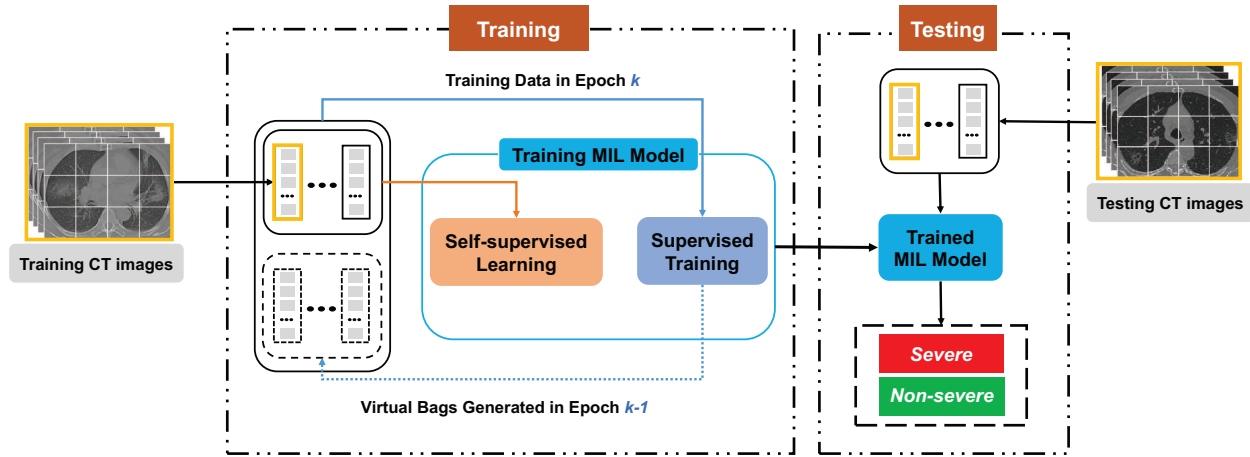}}
\caption{An overview of our method. In our method, the CT images are cropped into patches, which are then packed into MIL bags. In the $k^{th}$ epoch of training process, the data for supervised training consists of real bags (\ie, training CT images) and virtual bags generated in the $(k-1)^{th}$ epoch. Besides, real bags are also used for the auxiliary self-supervised learning task (while virtual bags are not). After the training stage, the trained MIL model will take the testing CT images (also modeled as MIL bags) as input to predict their labels (\ie, severe or non-severe).}
\label{ovv}
\end{figure*}

In this paper, we propose a method consisting of three major components (see Fig. \ref{ovv}). Specifically, 1) we build a deep multiple instance learning (MIL) model with instance-level attention to jointly classify the bag and weigh the instances in each bag, so as to find the positive key instances (\ie, instances with high confidence to the class of ``severe''); 2) We develop an instance-level augmentation technique to generate virtual positive bags by sampling from these key instances, which helps to ease the problem of class imbalance and strengthen the learning process; 3) We introduce an auxiliary self-supervised loss to render extracted features more discriminative, by including characteristic information of the patches. With extra information extracted from the unannotated instances, the  performance of MIL model could be further improved. \rev{These three components are logically integrated in a unified framework: 1) Three components are alternatively updated to benefit each other in an end-to-end manner; 2) Data augmentation could alleviate the label imbalance issue in training of the MIL model, while the trained MIL model could guide data augmentation to produce more meaningful bags; 3) Self-supervised pretext task is able to benefit the MIL model to being location-aware, which was ignored in traditional MIL setting. 
In our evaluation, we extensively validated the efficacy of our three components.}\par

In the following four sections, we first introduce related literature  (Section \ref{sec2}), then we present the technical details of the proposed method (Section \ref{sec3}), and finally report the qualitative and quantitative experimental results (Section \ref{sec4}) before drawing a conclusion (Section \ref{sec5}).

\section{Related Work}
\label{sec2}
We would like to review related work on four aspects: 1) COVID-19 severity assessment, 2) multiple instance learning, 3) data augmentation, and 4) self-supervised learning.

\subsection{COVID-19 Severity Assessment}
Along with diagnosis, severity assessment is another important factor for treatment planning. \rev{So far, there have been a few relevant attempts at predicting severity of COVID-19 with CT images.} Tang \etal \cite{ra78} proposed a random forest (RF)-based model to assess the severity of COVID-19 based on 63 quantitative features, \eg, the infection volume/ratio of the whole lung and the volume of ground-glass opacity (GGO) regions. Besides, importance of each quantitative feature is calculated from the RF model. Yang \etal \cite{yang2020chest} proposed to evaluate the value of chest computed tomography severity score (CT-SS) in differentiating clinical forms of COVID-19. The CT-SS was defined by summing up individual scores from 20 lung regions. In Yang \etalns's work, these scores were assigned for each region based on parenchymal opacification. Li \etal \rev{\cite{li2020ct} proposed a similar method depending on the total severity score (TSS), which was reached by summing the five lobe scores. Shan \etal \cite{shan2020abnormal} developed a deep learning based segmentation system, namely ``VB-Net", to automatically segment and quantify infection regions in CT scans of COVID-19 patients. To accelerate the manual delineation of CT scans for training, a human-involved-model-iterations (HIMI) strategy was adopted to assist radiologists to refine automatic annotation of each training case. Above existing methods all depended on manually defined scores or manual segmentations given by experts. Chao \etal \cite{chao2020integrative} and Chassagnon \etal \cite{chassagnon2020ai} further extended the problem to patient outcome prediction, combining both imaging and non-imaging (clinical and biological) data.} \par

\subsection{Multiple Instance Learning}
\rev{Multiple instance learning is one paradigm of weakly supervised learning, which belongs to ``inexact supervision" \cite{zhou2018brief}.} In terms of existing MIL methods for image classification task, they can be roughly divided into two categories, \re{\ie,} instance-level methods and bag-level methods. Instance-level methods assume that all instances in a bag \re{contribute equally} to the \re{prediction of} the bag label \cite{rb1}, under which assumption, the prediction of bag\re{-level} label is conducted by aggregating (\eg, voting or pooling) the prediction of instance\re{-level labels} in each bag. \re{However, this type of approaches \cite{rb23,rb30} suffer from a major limitation --} the label of each instance is usually separately predicted without considering other instances (even those in the same bag), rendering the label easily disrupted by incorrect instance-level predictions. However, \re{different from instance-level methods,} by considering the class information of all instances, bag-level methods usually achieve higher accuracy and better time efficiency, as proved in \cite{rb1}. In this sense, the promising property of MIL works quite well with our weakly supervised image classification task, as indicated by many current studies working towards this direction \cite{rb3,rb27,rb30}. \par
As is known, medical images are usually infeasible to obtain pixel-level annotations because this demands enormous time and perfect accuracy from clinical experts. Therefore, there has been a great interest in applying MIL methods to medical imaging \cite{cheplygina2019not}. Quellec \etal \cite{rbQ} attempted to divide the medical image into small-sized patches \re{that can} be considered as a bag with a single label. Sirinukunwattana \etal \cite{rbS} further extended this application to the computational histopathology where patches correspond to cells to indicate malignant changes. Ilse \etal \cite{admil} proposed an attention-based method that aims at incorporating interpretability to the MIL approach while increasing its flexibility \re{at the same time}. Han \etal \cite{ad3dmil} innovatively incorporated  an automated deep 3D instances generator into the attention-based MIL method, for accurate screening of COVID-19. There are also other different MIL approaches used in medical image analysis tasks, \eg, Gaussian processes \cite{rbK} and a two-stage approach with neural networks and expectation maximization (EM) algorithm to determine the classes of the instances \cite{rbH}. 

\subsection{Data Augmentation}
Data augmentation is a data-space solution to the problem of limited data. To increase the amount and the diversity of data, there has been a great interest in data augmentation recently, since many applications, \eg, medical image analysis, might not always have sufficient labeled training data \re{to train}. So far, a number of techniques have been developed to enhance the size and quality of training sets to build better deep learning models.\par 
One type of data augmentation methods are designed by performing the basic image processing operators. For example, Taylor and Nitschke \cite{rc63} provided a comparative study of the effectiveness of geometric transformations (\eg, flipping, rotating, and cropping), and that of color space transformations (\eg, color jittering, edge enhancement and PCA). Zhang \etal \cite{mixup} proposed \emph{mixup}, which trains the learning model on virtual examples constructed by a linear interpolation of two random examples from the training set. Zhong \etal \cite{rc70} developed random erasing inspired by the mechanisms of dropout regularization. Similarly, DeVries and Taylor \cite{rc71} proposed a method named as Cutout Regularization. \par
Note that there are also several attempts at learning-based data augmentation. Frid-Adar \etal \cite{rc49} tested the effectiveness of using DCGANs to generate liver lesion medical images. Applying meta learning concepts in neural architecture search (NAS) to data augmentation, several methods such as Neural Augmentation \cite{rc36}, Smart Augmentation \cite{rc37}, and Auto-Augment \cite{cubuk2019autoaugment}, were further developed in recent literature. \par
Unfortunately, under the MIL setting, the labels of instances are not available during training, which means previous data augmentation methods cannot be directly borrowed. \re{In order to relieve the data scarcity problem during COVID-19 severity assessment,} we have to develop a novel augmentation technique that works for our MIL setting.

\subsection{Self-supervised Learning}
In many recent studies of unsupervised learning, a common method is to define an annotation-free pretext task to provide a surrogate supervision signal for feature learning. By solving such pretext tasks, the trained model \re{is expected to} extract high-level semantic features that are useful for other downstream tasks. So far, a large number of pretext tasks for self-supervised learning have been designed. \re{For example}, Larsson \etal \cite{Larsson16} and Zhang \etal \cite{Zhang16} predicted the colors of images \re{by removing its original color information}. Doersch \etal \cite{Doersch15} and Noroozi and Favaro \cite{Noroozi16} predicted relative positions of \re{different} image patches \re{in the same image}. Gidaris \etal \cite{Gidaris18} predicted the random rotation applied to an image. Pathak \etal \cite{Pathak16} predicted the missing central part of an image \re{by building the prediction model with context information}. He \etal \rev{\cite{he2020momentum} presented a contrastive learning method, called Momentum Contrast (MoCo), which outperformed its supervised pre-training counterpart in several vision tasks. Zhou \etal \cite{ZHOU2021101840} proposed a set of models trained by a robust, scalable self-supervised learning framework, called Models Genesis, for medical image analysis tasks.} What these works have in common is that they are all utilized to attain well pre-trained networks on unannotated images. 

Unlike these works above aiming at pre-trained models, Chen \etal \cite{Chen18} aimed to improve the performance of generative adversarial networks by leveraging the supervision of rotation prediction task. Similarly, Gidaris \etal \cite{bf3s} used self-supervision as an auxiliary task and brought significant improvements to few-shot learning.

~\\
\noindent
\re{\textbf{Remark.} As discussed above, both the weak supervision manner and data scarcity in COVID-19 severity assessment pose a considerable challenge to our work. So our intuition includes the following two steps: 1) We found weak supervised prediction of COVID-19 naturally agrees with the setting of MIL; 2) Under the MIL setting, we try to solve the challenge of data scarcity by considering the relation between bag and instance. In a nutshell, by confronting the double challenges of weak supervision and data scarcity, our solution for COVID-19 severity assessment is novel according to our best knowledge.}

\section{Method}
\label{sec3}
In this section, we first analyze the problem of COVID-19 severity assessment and provide an overview of our method, then present and discuss thoroughly the technical details of three major components, \ie, bag-level prediction, instance-level augmentation and auxiliary self-supervised loss.

\subsection{Problem Analysis}
In this part, we will first analyze the main challenges in COVID-19 severity assessment caused by \re{weak supervision} and data scarcity, and then provide corresponding countermeasures.  

For the annotation \re{of CT images}, \rev{image-level labels directly come from diagnosis results of corresponding patients, guided by the \textit{Diagnosis and Treatment Protocol for COVID-19 (Trial Version 7)} from National Health Commission of the People's Republic of China.} \re{In this sense}, infection regions in CT images \re{of COVID-19 patients} remain unknown even when this image has already been labeled. This poses a great challenge for the utilization of traditional supervised learning models. To address this challenge, we introduce the multiple instance learning (MIL) framework, a typical weakly-supervised learning \re{paradigm} to deal with the image-level prediction \re{without knowing any region-level annotation}.

\begin{figure}[htbp]
\centerline{\includegraphics[scale=.3]{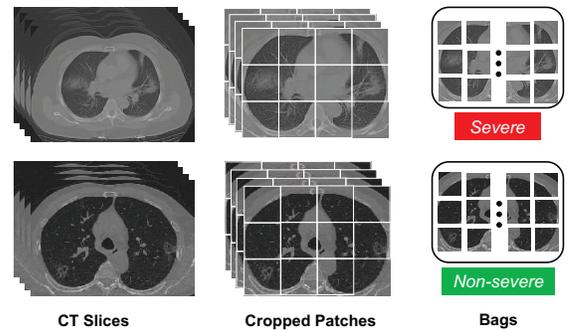}}
\caption{A brief illustration of the notions of CT slices, instances (patches) and bags. A CT image contains CT slices, and the slices are cropped to non-overlapping patches, which are considered as instances. The patches from the same CT image make up a MIL bag, with a bag-level label ``severe'' or ``non-severe''.}
\label{notions}
\end{figure}

In the MIL setting, each image could be regarded as a bag, and regions inside this image are thus regarded as instances in this bag. In our case, chest CT images are processed as bags. To be more specific, each CT image consists of hundreds of slices that show different cross sections of lung regions. Moreover, each slice can be further cropped into several \re{non-overlapping} patches. And the patches from the same CT image make up a bag. Note that the label of a bag (\ie, the bag-level label) depends on \re{the information provided by physicians on} original CT images. These notions are illustrated in Fig. \ref{notions}.  In this work, the MIL bags with the label ``severe'' are called \emph{positive bags}, whereas those with the label ``non-severe'' are called \emph{negative bags}. \re{It is also worth mentioning that the instances (\ie, patches) related to the infection regions in positive bags are without any annotated information} during training. 

Another challenge is data scarcity, which usually makes stable learning model hard to \re{realize}, especially when the number of severe CT images is very limited. To address this issue, we first adopt the data augmentation technique during training by generating virtual samples. Though data augmentation has demonstrated its effectiveness in several other learning tasks \cite{shorten2019survey,qin2020unsupervised}, previous augmentation technique cannot be directly applied to our MIL setting, because each bag consists of several instances and the instance-level label is unknown. In our work, we notice that compared to other instances, some instances usually play a much more important role in determining the label of a bag, \re{which we name as \emph{key instances}.} This observation further drives us to develop a novel instance-level augmentation technique to generate ``virtual'' bags by gathering these key instances.

In addition to the data augmentation technique, we also leverage self-supervised learning, a popular unsupervised paradigm in recent studies. We notice that, although supervised information of bag-level labels is limited due to data scarcity, there is a wealth of unsupervised information hidden in unannotated instances. \re{Thanks to} self-supervision, \re{the patch-wise location can be further exploited} so that the network can extract stronger instance features. As a result, the bag-level features of positive and negative samples can be further differentiated, further improving the performance of the MIL model.

In a word, to address the aforementioned challenges, our method has three major components: 1) bag-level prediction, 2) instance-level augmentation and 3) auxiliary loss based on self-supervised learning. In particular, we build a deep multiple instance learning model with attention mechanism to predict the bag-level label (\ie, severe or non-severe). Instances with higher attention weights, that are expected to have larger influence on the bag-level label, can be regarded as key instances, while other instances are considered to be regular instances. According to the learned attention, we randomly sample key instances and regular instances to generate virtual bags to enrich current training samples. In the training stage, we incorporate self-supervision into MIL model by adding an auxiliary self-supervised loss.

\begin{figure*}[htbp]
\centerline{\includegraphics[scale=.45]{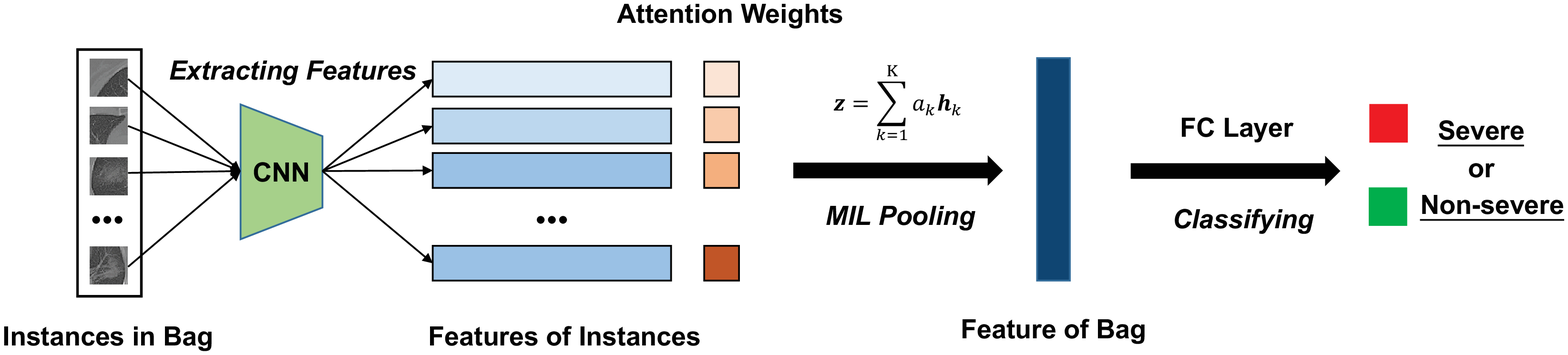}}
\caption{The framework of deep MIL model. Firstly, the instance features are extracted. Secondly, the attention weights of the instance features are determined by the network. Then, the MIL pooling layer combines the instance features to generate a bag feature. Finally, the bag feature is mapped by a fully connected (FC) layer to decide the label.}
\label{mil}
\end{figure*}

\subsection{Bag-level Prediction}
The first component of our method -- bag-level prediction aims to predict the label of a CT image as either severe or non-severe. A chest CT image consisting of hundreds of slices can be divided into smaller sized patches, making the CT image itself a bag with a single label (severe or non-severe) and the patches instances. In this work, $Y=1$ indicates that the image is labeled as severe case while $Y=0$ indicates that it is non-severe. \re{Since these patches are non-overlapping,} we assume there is no dependency or sequential relationship among the instances within a bag. Furthermore, $K$ denotes the number of instances in a bag, and we assume $K$ could vary from bag to bag. \par

The framework of this component is shown in Fig. \ref{mil}. We use a convolutional neural network \cite{lenet} to extract the feature embedding $\mathbf{h}_{k}$ of each instance $\mathbf{x}_{k}$, where $\mathbf{h}_k \in \mathbb{R}^{M}$ and $M$ is the dimensionality of instance features. Suppose $H=\left\{\mathbf{h}_{1}, \ldots, \mathbf{h}_{K}\right\}$ is a bag of $K$ embeddings, the embedding of bag $X$ is calculated by attention-based MIL pooling \rev{proposed by Ilse \etal \cite{admil}}:
\begin{equation}
\mathrm{z}=\sum_{k=1}^{K} a_{k} \mathbf{h}_{k},
\end{equation}
where
\begin{equation}
a_{k}=\frac{\exp \Big(\mathbf{w}^{\top} \tanh \left(\mathbf{V h}_{k}^{\top}\right) \Big)}{\sum_{j=1}^{K} \exp \Big(\mathbf{w}^{\top} \tanh \left(\mathbf{V h}_{j}^{\top}\right)\Big)},
\end{equation}
$\mathbf{w} \in \mathbb{R}^{L}$ and $\mathbf{V} \in \mathbb{R}^{L \times M}$ are the parameters to learn. \rev{The hyperbolic tangent $\tanh(\cdot)$ element-wise non-linearity is utilized to include both negative and positive values for proper gradient ﬂow.} Finally, we use a fully connected layer to decide the label according to the bag feature. The categorical cross-entropy loss is used to optimize the MIL model deﬁned as follows:
\begin{equation}
L_\text{MIL}=-\frac{1}{N_b} \sum_{i=1}^{N_b} \sum_{c=0}^{1} \delta\left(Y_{i}=c\right) \log \Big(P\left(Y_{i}=c\right)\Big),
\end{equation}
where $N_b$ is the number of bags. $\delta\left(y_{i}=c\right)$ is the indicator function \re{(\ie, $\delta\left(y_{i}=c\right)=1$ when $y_{i}=c$ and 0 otherwise)} and $P(Y_i = c)$ denotes the predicted probability.\par 
It is worth mentioning that large weights refer to key instances with relatively high confidence, that are most relevant to the bag-level label. \re{This means that} not only can the MIL model provide final diagnostic results, it can also help physicians to identify possible \textbf{severe} infection regions, \re{which has a great clinical significance for COVID-19 severity assessment.}\par 
With the trained MIL model, we are able to automatically assess the severity of the disease with CT images. In the testing stages, we divide the CT image \re{into non-overlapping} patches in the same way as in training. Along with the assessment, the model also outputs the attention weights, that can help find the regions relevant to severe infection.

\subsection{Instance-level Augmentation}
For severity assessment, scare data significantly deteriorates the overall performance. What's more, the imbalance of class, \ie, the number of non-severe cases is much larger than that of severe cases, is also harmful for learning a stable model. \par
\begin{figure}[htbp]
\centerline{\includegraphics[scale=.4]{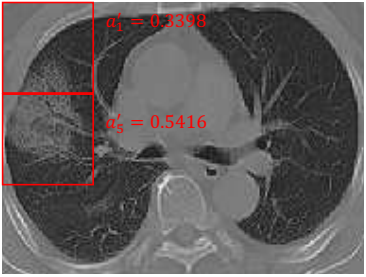}}
\caption{An example of key instances in a positive bag. It indicates that the patches are likely to be related to the severe infection regions. Note that, we rescaled the attention weights of the patches in the same slice using $a_k^{'} = a_k / \sum_i a_i$}
\label{key}
\end{figure}

To confront these problems, we propose a novel data augmentation technique to generate ``virtual bags'' on original bags to enrich current training process. By rethinking the attention mechanism in multiple instance learning, we notice that the patch with higher responses in attention usually indicates a higher relation to its current class label. In positive bags, there are some instances with significantly higher weights, as shown in Fig. \ref{key}. We consider them \emph{(positive) key instances} while other instances \emph{regular instances}. However, according to the experiments, in negative bags, all instances have similar low weights, \re{roughly confirming the rule in traditional MIL}. Therefore, we only take positive bags and corresponding key instances in them into consideration.

\begin{figure}[htbp]
\centerline{\includegraphics[scale=.37]{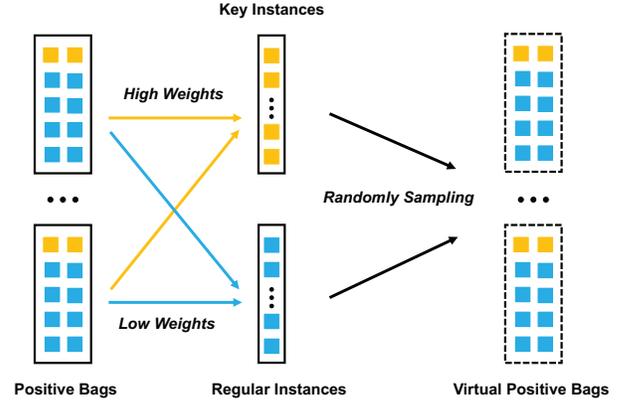}}
\caption{A sketch map of generating virtual bags. For positive bags, instances with high weights are appended to the list of key instances while instances with low weights to the list of regular instances. Virtual bags are generated by randomly sampling key instances and regular instances.}
\label{fig1}
\end{figure}

In the training process, when a positive bag is correctly predicted \re{to be severe}, instances with top-$\lfloor \alpha K \rfloor$ highest weights \re{in this bag} will be appended to the list of key instances, where $K$ is the number of instances in the bag. In the meantime, other instances are treated as regular instances. Considering time \re{cost} and memory usage, we only append instances with top-$\lfloor \gamma N \rfloor$ lowest weights to the list of regular instances. In our observation, we notice that different positive bags may have different proportion of positive key instances. Therefore, in order to prevent regular instances from being mistaken \re{as} key instances, the proper parameter $\alpha$ should be set to a relatively small value \re{as a strict rule to judge the key instance. In practice, the value is set to} no larger than each positive bag's actual proportion. For the generation of virtual positive bags, assuming the average number of instances per bag is $\bar{K}$, we randomly sample $\lfloor \alpha \bar{K} \rfloor$ key instances and $\lfloor (1-\alpha)\bar{K} \rfloor$ regular instances, then pack them into a \re{virtual} bag. Among the parameters above, $K$ and $\bar{K}$ are easy to obtain, while $\alpha$ and $\gamma$ need to be set before training. The process of generating virtual bags is visualized in Fig. \ref{fig1}.\par

With the help of attention mechanism, this process can be plugged in the training phase. In \re{each} training epoch, the model generates virtual bags based on attention weights and these virtual bags generated \re{are further} used as a part of training data in the next epoch. \par  

\subsection{Auxiliary Self-supervised Loss}
We  also incorporate self-supervision into the MIL model by adding an auxiliary self-supervised loss. By optimizing the self-supervised loss, the model learns to exploit more information from unannotated instances.\par
In this work, we consider the following two pretext tasks for the self-supervised loss: 1) \re{to predict} the relative location of two patches from the same slice, which is a seminal task in self-supervised learning \re{as originally proposed in} \cite{Doersch15}; and 2) \re{to predict} the absolute location of a single patch, which is similar to the former task, but more suitable under the MIL problem setting.

\begin{figure}[htbp]
\centerline{\includegraphics[scale=.4]{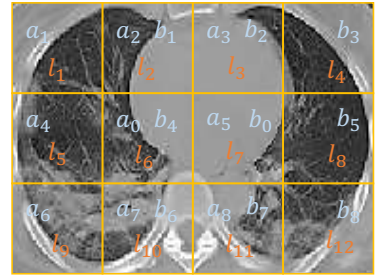}}
\caption{A CT slice is divided to 12 patches. For the relative patch location task, we are able to create 16 pairs of patches: $(a_0, a_1), \dots ,(a_0,a_8)$ and $(b_0,b_1), \dots ,(b_0,b_8)$. For the absolute patch location task, we directly predict each patch's location among $l_1,\dots l_{12}$.}
\label{patches}
\end{figure}
\noindent
\textbf{SSL task 1: Relative Location Prediction.} Predicting the relative location of a pair of patches from the same image is a seminal task in self-supervised learning. More specifically, given a pair of patches, the task is to predict the location of the second patch with regard to the first one, among eight possible positions, \eg, ``on the bottom left'' or ``on the top right''. This task is particularly suitable for the MIL setting, because there are many pairs of patches in a bag, that we can predict the relative location of. To be more specific, for one slice, we are able to create 16 pairs of patches: $(a_0, a_1), \dots, (a_0,a_8)$ and $(b_0,b_1), \dots, (b_0,b_8)$, as shown in Fig. \ref{patches}. We extract the representation of each patch and then generate pair features by concatenation. We train a fully connected network $\mathbf{G}_{\phi}^r(\cdot, \cdot)$ with parameters $\phi$ to predict the relative patch location of each pair. \par
The self-supervised loss of \re{this relative location prediction} task is defined as:
\begin{equation}
    L^r_\text{SSL} = \frac{1}{N_s} \sum_{s=1}^{N_s} \sum_{i=1}^{16} L_\text{CE}\Big(\mathbf{G}_\phi^r(p_i), \texttt{rloc}(p_i)\Big),
\end{equation}
where $N_s$ is the number of slices. $p_i$ stands for the pair of patches, specifically $p_1, \dots, p_8$ for  $(a_0, a_1), \dots ,(a_0,a_8)$ and $p_9, \dots ,p_{16}$ for $(b_0,b_1), \dots ,(b_0,b_8)$. There are 16 pairs in total. $L_\text{CE}(\cdot,\cdot)$ is the cross-entropy loss function and $\texttt{rloc}(p_i)$ is the ground truth of the relative patch location. \par
~\\
\noindent
\textbf{SSL Task 2: Absolute Location Prediction.} 
Under the MIL setting of COVID-19 severity assessment, the CT slices containing two lungs are quite similar. \re{This has made us to realize that} the task can be designed in a more straightforward way, \ie, to predict the absolute location of a single patch \re{in an entire CT slice},  or more specifically,  to predict the location of a patch among 12 possible positions $l_1,\dots l_{12}$ as shown in Fig. \ref{patches}. We also train a fully-connected network $\mathbf{G}_{\phi}^a(\cdot)$ with parameters $\phi$ to predict the absolute patch location.\par
The self-supervised loss of \re{this absolute location prediction} task is defined as:
\begin{equation}
    L^a_\text{SSL} = \frac{1}{N_s} \sum_{s=1}^{N_s} \sum_{i=1}^{12} L_\text{CE}\Big(\mathbf{G}_\phi^a(x_i), \texttt{aloc}(x_i)\Big),
\end{equation}
where $N_s$ is the number of slices. $x_i$ stands for the patch in the position of $l_i$. $\texttt{aloc}(x_i) = l_i$ is the ground truth of the absolute patch location. There are 12 patches per slice.

~\par
Formally, let $L_\text{SSL}$ be the either kind of self-supervised loss, the total loss of the training stage can be written as
\begin{equation}
    L_\text{total} = \frac{L_\text{MIL} + \mu L_\text{SSL}}{1 + \mu}, 
\end{equation}
where $L_\text{MIL}$ stands for the loss of MIL model (\ie, the bag-level prediction task), as defined previously. The positive hyperparameter $\mu$ controls the weight of self-supervised loss. By optimizing self-supervised loss, the instance feature extractor can learn more informative features, further improving the performance of the MIL model. Note that, only real bags will be used for self-supervised learning tasks.

\section{Experiment}
\label{sec4}
In the section, we will report the quantitative and qualitative results of our method. First, we present the details of our COVID-19 dataset and the process of data preprocessing. Then, we discuss the experimental setup in our evaluation and provide the implementation details of our method. After that, we conduct ablation studies, and compare our method with existing methods, while analyzing the interpretability of our method. Finally, we discuss some of the choices we have made in network structures, parameter values and pretext tasks.

\subsection{Dataset}
\begin{figure}[htbp]
\centerline{\includegraphics[scale=.5]{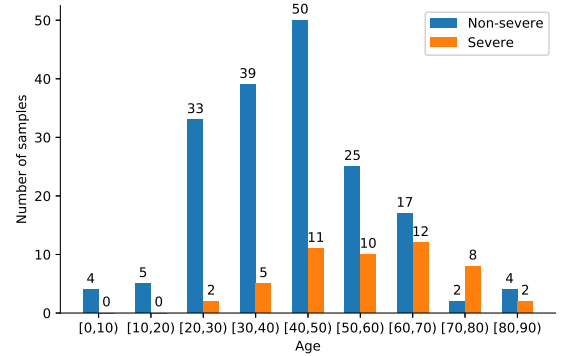}}
\setlength{\abovecaptionskip}{-0.2cm}  
\setlength{\belowcaptionskip}{-0.5cm}
\caption{The age distribution of the patients in our dataset.}
\label{age}
\end{figure}

We collect a dataset, that contains chest CT images of 229 patients with confirmed COVID-19. \rev{For each patient, the severity of COVID-19 is determined according to \textit{Diagnosis and Treatment Protocol for COVID-19 (Trial Version 7)} issued by National Health Commission of the People's Republic of China.} The severity includes four types: mild, common, severe and critical. We categorize patients into two groups: non-severe group (mild and common) and severe group (severe and critical), because the number of patients with mild or critical types is extremely small. Among these patients, 179 are non-severe cases while 50 severe cases. \rev{The categories of patients are used as image-level labels of their corresponding CT images.} Moreover, the gender distribution of the patients is shown in Table \ref{gender} and their age distribution is shown in Fig. \ref{age}.\par 

\begin{table}[htbp]
    \renewcommand\arraystretch{1.1}
  \centering
  \caption{The Gender Distribution in our study}
  \label{gender}
    \begin{tabularx}{8.3cm}{|p{40pt}<{\centering}|p{50pt}<{\centering}p{50pt}<{\centering}|X<{\centering}|}
    \hline
    Gender & Severe & Non-severe & Total \\
    \hline
    \hline
    Male  & 32    & 92    & 124 \\
    Female & 18    & 87    & 105 \\
    \hline
    \hline
    Total & 50    & 179   & 229 \\
    \hline
    \end{tabularx}
\end{table}%

All chest CT images were acquired at the Second Xiangya Hospital of Central South University and its collaborating hospitals with different types of CT scanners, including Anke (ANATOM 16HD), GE (Bright speed S), Hitachi (ECLOS), Philips (Ingenuity CT iDOSE4) and Siemens (Somatom perspective). The scanning parameters are as follows: 120 kVp, 100-200 mAs, pitch 0.75-1.5 and collimation 1-5 mm. 

\subsection{Data Preprocessing}
Though in the previous section, we have introduced how to process the CT images as MIL bags briefly, when it comes to the implementation, we give more details. \rev{The CT images are originally stored in DICOM files and converted to PNG file format for further processing. Each PNG file corresponds to a CT slice. After the CT images have been sliced, those slices with few lung tissues are removed. For each slice, we locate the bounding box of two lungs and crop the region containing two lungs. The cropping region is then resized to $240\times 180$ and divided into 12 non-overlapping patches of size $60\times 60$. To remove inter-subject variation, we further perform min-max normalization on each patch individually.}

Eventually, we obtained a dataset consisting of 229 bags, 179 of which are negative (\ie, non-severe cases) and 50 of which are positive (\ie, severe cases). There are 49,632 instances (patches) in total, with around 217 instances per bag on average.

\subsection{Experimental Setup}
\rev{We employ the standard 10-fold cross-validation, where each sample would be tested at least once. Each experiment is performed 5 times and an average ($\pm$ a standard deviation) is reported to avoid possible data-split bias.} 

In our experiments, we use the following metrics to evaluate the performance: accuracy, sensitivity (true positive rate, TPR), specificity (true negative rate, TNR), F1-Score and the area under the receiver operating characteristic curve (AUC). Specifically, these metrics are defined as:
\begin{equation}
\text {Accuracy}=\frac{\text{TP}+\text{TN}}{\text{TP}+\text{FP}+\text{FN}+\text{TN}},
\end{equation}
\begin{spacing}{0.75}
\begin{equation}
\text{Sensitivity} = \frac{\text{TP}}{\text{TP}+\text{FN}},
\end{equation}

\begin{equation}
\text{Specificity} = \frac{\text{TN}}{\text{TN}+\text{FP}},  
\end{equation}

\begin{equation}
\text {F1-score }=\frac{\text{TP}}{\text{TP} + \frac{1}{2}(\text{FP} + \text{FN})},
\end{equation}
\end{spacing}
\vspace{1ex}
\noindent where TP, FP, TN, and FN represent the True Positive, False Positive, True Negative and False Negative, respectively.
\subsection{Implementation Details}
In this part, we provide the implementation details of the three components of our method.
\subsubsection{Bag-level Prediction}
For bag-level prediction, we construct a deep attention-based MIL model. In order to keep the consistency with the previous work \cite{admil}, we choose LeNet \cite{lenet} as the instance feature extractor and the dimensionality of features is 512. In attention mechanism, the parameter $L$ is set as 128. A fully connected (FC) layer with Sigmoid function works as a linear classifier and the classification threshold is set as 0.5. All layers are initialized according to \cite{gb} and biases are set as zero. The network architecture is shown in Table \ref{tab1}. 
In Table \ref{tab1}, conv (5,1,0)-36 means the size of kernel as 5, stride as 1, padding as 0 and the number of output channels as 36, respectively.
The model is trained with the Adam optimization algorithm \cite{adam}. The hyperparameters of the optimization procedure are given in Table \ref{tab2}.\par
\begin{table}[htbp]
\centering
\caption{The details of our Network Architecture}
\label{NA}
\renewcommand\arraystretch{1.1}
\begin{tabularx}{7.8cm}{|p{60pt}<{\centering}|X<{\centering}|}
\hline
Layer & Type \\
\hline
\hline
1 & conv(5,1,0)-36 + ReLU \\
2 & maxpool (2,2) \\
3 & conv(5,1,0)-36 + ReLU \\
4 & maxpool (2,2) \\
5 & conv(5,1,0)-48 + ReLU \\
6 & maxpool (2,2) \\
7 & fc-512 + ReLU \\
8 & MIL-attention-128\\
9 & fc-1 + Sigm \\
\hline
\end{tabularx}
\label{tab1}
\end{table}

\begin{table}[htbp]
\centering
\caption{The setting of parameters in our experiments}
\label{tab2}
\renewcommand\arraystretch{1.1}
\begin{tabularx}{6.8cm}{|p{80pt}<{\centering}|X<{\centering}|}
\hline
Hyperparameters & Value \\
\hline
\hline
$\beta_1,\beta_2$ & 0.9, 0.999 \\
Learning rate & 0.0001 \\
Weight decay & 0.0005 \\
Batch Size & 1 \\
Epoch & 50 \\
\hline
\end{tabularx}
\end{table}

\subsubsection{Instance-level Augmentation}
In virtual-bag generation, the value of parameter $\alpha$ is very important. As aforementioned in Section \uppercase\expandafter{\romannumeral4}.C, in our setting, $\alpha$ should be smaller than the actual proportion to prevent regular instances from being mistaken for key instances. So we set $\alpha$ as 0.025, because it shows the greatest accuracy on the validation set. On the contrary, since regular instances are not useful to identify the positive bag, the parameter $\gamma$ just need to be relatively small, to make sure that no key instance gets into the list of regular instances incorrectly. In our experiments, $\gamma$ is fixed as 0.2, which shows good performance according to our evaluation. Besides, in the beginning of the training phase, the MIL model is under-fitting so it cannot provide very accurate weights. Therefore, in our experiment, the model starts to generate virtual bags from the $26^{th}$ epoch to the last epoch.\par

\subsubsection{Self-supervised Loss} For the relative patch location task, given two patches, the network $\mathbf{G}_{\phi}^r(\cdot)$ gets the concatenation of their feature vectors as input to two fully connected layers. For the absolute patch location task, another network $\mathbf{G}_\phi^a(\cdot)$ consisting of two fully connected layers gets as input the feature of a single patch and predicts its location. For both tasks, we set $\mu$ as 0.3.
We also optimize the auxiliary loss with the Adam algorithm \cite{adam}. The hyperparameters of the optimization algorithm are consistent with those shown in Table \ref{tab2}.\par

\subsection{Ablation Study}
To evaluate the effectiveness of different components of the proposed method, we have conducted ablation studies. We have experimented on the following configurations: 

\begin{itemize}
\setlength{\itemsep}{0pt}
\setlength{\parsep}{0pt}
\setlength{\parskip}{0pt}
\item (A) MIL Only: our method without data augmentation and self-supervised learning;
\item (B) MIL + Augmentation: our method without self-supervised learning; 
\item (C) MIL + Self-supervised: our method without data augmentation;
\item (D) MIL + Both: \textbf{our method}.
\end{itemize}
The values of hyperparameters involved have already been described above. For self-supervised learning, we choose the absolute patch location task as the pretext task, because experimental evidence shows that it outperforms the relative patch location task in our problem setting. See further discussion for details. \par

The results of all these conﬁgurations are illustrated in Table \ref{ablation}. \rev{The statistical comparison (\ie, two-sample t-test) on AUC metrics is conducted and p-values are reported below.} Comparing (B) with (A), we find that the proposed data augmentation technique significantly improves the overall performance of the MIL model \rev{($p=0.015<0.05$)}, especially the sensitivity criteria important for COVID-19 diagnosis. Similarly, the comparison between (C) and (A) shows the auxiliary self-supervision loss also results in performance gain \rev{($p=0.028<0.05$)}. Comparing (D) with other configurations (A, B and C), the MIL model incorporating both data augmentation and self-supervised learning achieves the best performance \rev{($p=0.006, 0.027, 0.007<0.05$)}.\par

\begin{figure}[htbp]
\centerline{\includegraphics[scale=.25]{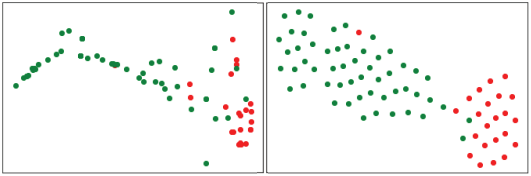}}
\caption{Visualization of the bag-level features extracted in different configurations. The left corresponds to (A), while the right corresponds to (D). Red and green points stand for severe and non-severe cases, respectively.}
\label{tsne}
\end{figure}

The visualization in Fig. \ref{tsne} indicates that our method can learn more discriminative feature representations than the original MIL model.

\begin{table*}[htbp]
\renewcommand\arraystretch{1}
  \centering
  \setlength{\belowcaptionskip}{10pt}
  \caption{The Results of The Ablation Study}
  \label{ablation}
    \begin{tabularx}{17cm}{p{115pt}<{\centering}X<{\centering}X<{\centering}X<{\centering}X<{\centering}X<{\centering}}
    \toprule
    Method & Accuracy & Sensitivity   &  Specificity & F1-Score & AUC \\
    \midrule
    (A) MIL Only  & \rev{$0.908 \pm 0.029$}    & \rev{$0.740 \pm 0.084$} & \rev{$0.954 \pm 0.031$} & \rev{$0.776 \pm 0.064$} & \rev{$0.932 \pm 0.022$}\\
    (B) MIL + Augmentation & \rev{$0.942 \pm 0.025$}    & \rev{$0.916 \pm 0.022$} & \rev{$0.949 \pm 0.036$} & \rev{$0.874 \pm 0.045$} & \rev{$0.971 \pm 0.004$} \\
    (C) MIL + Self-supervision & \rev{$0.937 \pm 0.029$}    & \rev{$0.844 \pm 0.129$} & \rev{$0.963 \pm 0.009$} & \rev{$0.850 \pm 0.079$} & \rev{$0.964 \pm 0.030$}\\
    (D) MIL + Both & \rev{\textbf{0.958 $\pm$ 0.015}}  & \textbf{\rev{0.936 $\pm$ 0.032}} & \textbf{\rev{0.964 $\pm$ 0.024}} & \textbf{\rev{0.895 $\pm$ 0.029}} & \textbf{\rev{0.981 $\pm$ 0.006}} \\
    \bottomrule
    \end{tabularx}%
\end{table*}%

\subsection{Comparison with Existing Methods}
We have compared our method with the existing works by Tang \etalns's \cite{ra78} and Yang \etalns's \cite{yang2020chest}, \rev{which share the same problem setting with ours}. For \rev{the size of} datasets, compared with chest CT images of 179 patients in Tang \etal's work and those of 102 patients in Yang \etal 's work, our work includes a larger dataset of 229 patients. \rev{In terms of data annotation, our method works under weak annotation setting, with only image-level labels (severe or non-severe) available. However, their works need additional manual annotation besides image-level labels. Tang \etal 's work depends on 63 quantitative features calculated from accurate segmentation results of infection regions. The segmentation network needs manual delineations for training. Yang \etal 's work depends on manually defined severity scores of lung regions provided by chest radiologists.}\par
Table \ref{tab3b} displays the comparison between our proposed method and the existing methods. Because their data and codes are not accessible, the results in the first two lines are directly reported from their papers. The third line shows that our MIL model itself has achieved better performance in terms of accuracy and AUC metrics. As shown in the last line, our proposed method with data augmentation and self-supervised learning reveals a superior performance on a larger dataset when compared with these two state-of-the-arts methods.\par

\begin{table*}[htbp]
\renewcommand\arraystretch{1}
  \centering
  \setlength{\belowcaptionskip}{10pt}
  \caption{Comparisons between Our Proposed Method and The Existing Approaches. Note that our method requires no addition annotation other than image-level labels, while Tang \etalns's work requires accurate segmentation of infection region and Yang \etalns's work depends on manually defined severity scores.}
  \label{tab3b}
    \begin{tabularx}{17cm}{p{110pt}<{\centering}X<{\centering}X<{\centering}X<{\centering}X<{\centering}X<{\centering}X<{\centering}}
    \toprule
    Method & Accuracy & Sensitivity   &  Specificity & F1-Score & AUC \\
    \midrule
    Tang \etalns's \cite{ra78}  & 0.875  & 0.933 & 0.745 & - & 0.910 \\
    Yang \etalns's \cite{yang2020chest}  & 0.833 & 0.940 & - & 0.892\\
    \midrule
    MIL Only (Ours) & \rev{$0.908 \pm 0.029$}    & \rev{$0.740 \pm 0.084$} & \rev{$0.954 \pm 0.031$} & \rev{$0.776 \pm 0.064$} & \rev{$0.932 \pm 0.022$}\\
    MIL + Both (Ours)  & \rev{\textbf{0.958 $\pm$ 0.015}}  & \textbf{\rev{0.936 $\pm$ 0.032}} & \textbf{\rev{0.964 $\pm$ 0.024}} & \textbf{\rev{0.895 $\pm$ 0.029}} & \textbf{\rev{0.981 $\pm$ 0.006}} \\
    \bottomrule
    \end{tabularx}%
\end{table*}%

\rev{
\subsection{Efficiency of MIL Method}
We implemented our experiments on one Nvidia GeForce RTX 2080 Ti GPU. In 10-fold cross-validation, for each data split, training our MIL model (with data augmentation and auxiliary self-supervised loss) on 206 samples would take $213.7 \pm 5.1$ seconds in average, and testing on 23 samples would take less than 1 second. It is shown that the proposed method is quite efficient in computation.}

\subsection{Interpretability of MIL Method}
\begin{figure*}[htbp]
\centerline{\includegraphics[scale=.45]{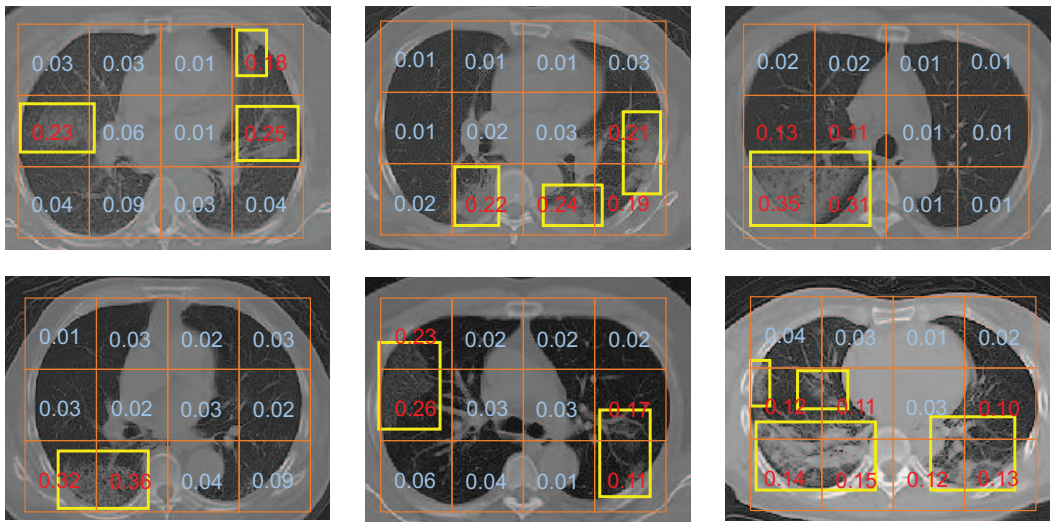}}
\caption{Visualization of the attention mechanism in our method. Presented above are some examples of the (slice-wise rescaled) attention weights of patches. Severe infection regions identified by experts are marked with yellow boxes. It can be seen that the patches with high weights are probably relevant to severe infection.}
\label{att}
\end{figure*}

Along with a predicted label, the MIL model also outputs attention weights for each patch. Although the model is not designed to accurately segment the lesions, it can still help to indicate the regions relevant to severe infection. In Fig. \ref{att}, we show that the attention weights can be useful for finding severe infection regions.  

\begin{table*}[htbp]
\renewcommand\arraystretch{1}
  \centering
  \setlength{\belowcaptionskip}{10pt}
  \caption{Performance of Different Pretext Tasks}
  \label{ptasks}
    \begin{tabularx}{17cm}{p{110pt}<{\centering}X<{\centering}X<{\centering}X<{\centering}X<{\centering}X<{\centering}}
    \toprule
    Method & Accuracy & Sensitivity   &  Specificity  & F1-Score & AUC \\
    \midrule
    Baseline & \rev{$0.908 \pm 0.029$}    & \rev{$0.740 \pm 0.084$} & \rev{$0.954 \pm 0.031$} & \rev{$0.776 \pm 0.064$} & \rev{$0.932 \pm 0.022$}\\
    Relative Patch Location  & \rev{$0.916 \pm 0.004$}    & \rev{$0.796 \pm 0.114$} & \rev{$0.949 \pm 0.032$} & \rev{$0.801 \pm 0.024$} & \rev{$0.951 \pm 0.014$}\\
    Absolute Patch Location  & \rev{\textbf{0.937 $\pm$ 0.029}}    & \rev{\textbf{0.844 $\pm$ 0.129}} & \rev{\textbf{0.963 $\pm$ 0.009}} & \rev{\textbf{0.850 $\pm$ 0.079}} & \rev{\textbf{0.964 $\pm$ 0.030}}\\
    \bottomrule
    \end{tabularx}%
\end{table*}%

\subsection{Method Designing Details}
Now we would like to discuss some of the choices we made in designing our method. Experiments on the validation set show that the values of parameters $\alpha$ and $\mu$, as well as the selection of pretext task, can affect the performance of our method.\par
\subsubsection{Bag-level Prediction} For the instance feature extractor, we do not use deeper ResNet \cite{resnet} or DenseNet \cite{densenet}, because the experiment shows that deeper networks cannot significantly improve the performance, but rather increase the time consumed instead. For MIL attention pooling, we test the following dimensions (L): 64, 128 and 256. The differences in dimensions only result in minor changes of the model’s performance.\par
\subsubsection{Instance-level Augmentation} 
\begin{figure}[htbp]
\setlength{\belowcaptionskip}{-0.3cm}
\centerline{\includegraphics[scale=.45]{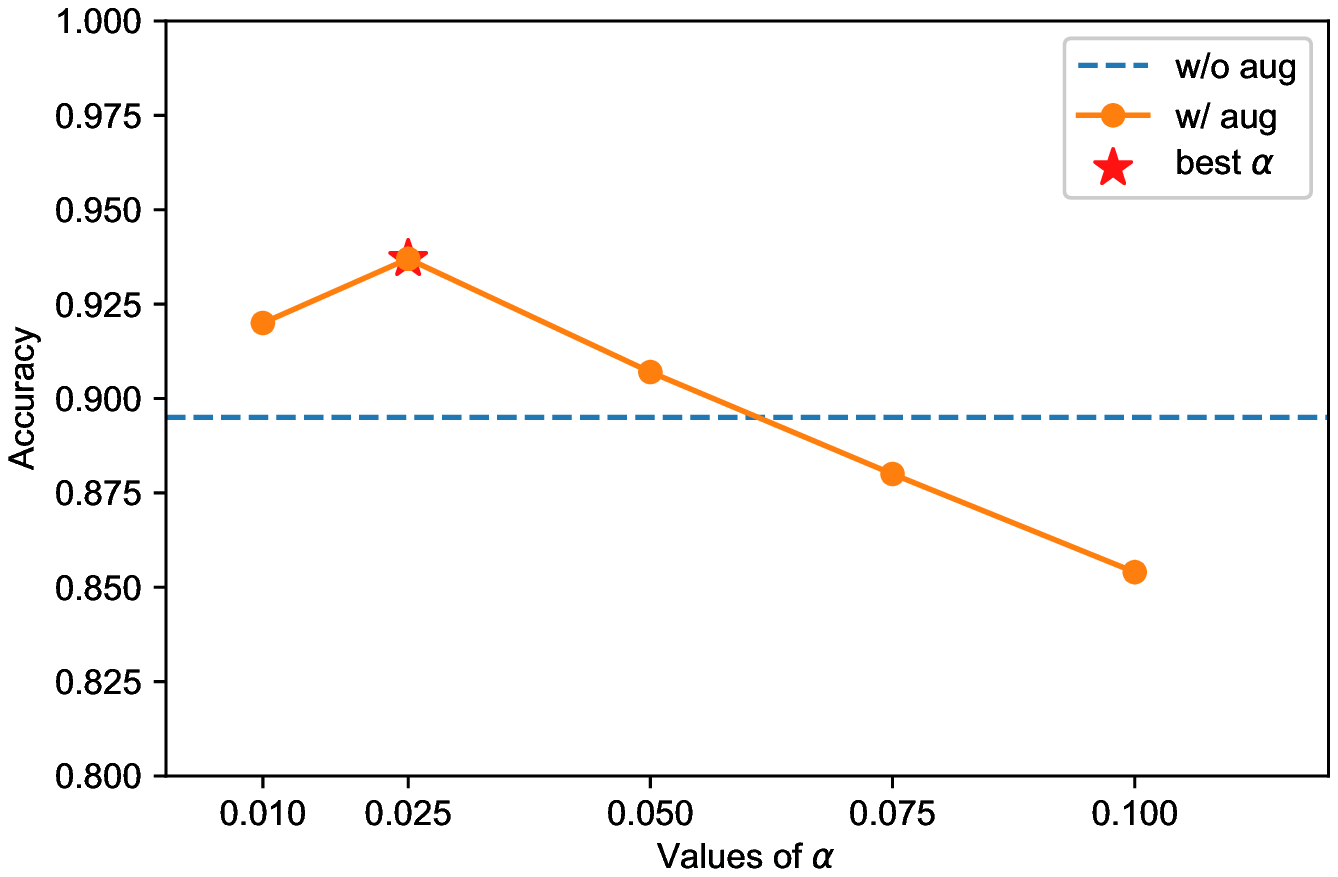}}
\caption{Performance of different $\alpha$. The best $\alpha$ is 0.025.}
\label{alpha}
\end{figure}

For the data augmentation technique, we evaluate the performance for different $\alpha$, the results of which are illustrated in Fig. \ref{alpha}. Our experiment shows that different $\gamma$ doesn't bring great variation on the model's performance.\par

\begin{figure}[htbp]
\setlength{\belowcaptionskip}{-0.5cm}  
\centerline{\includegraphics[scale=.45]{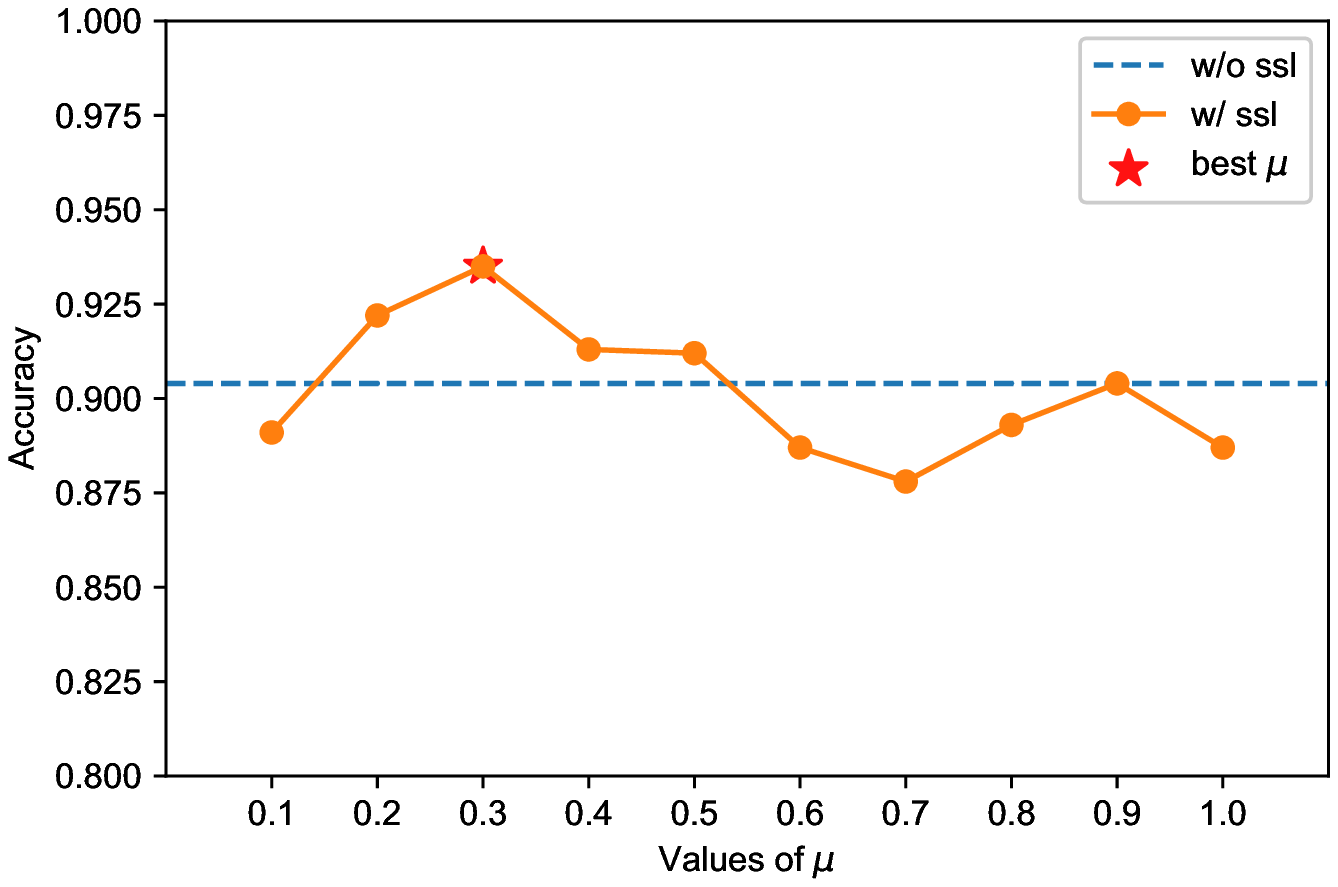}}
\caption{Performance of different $\mu$. The best $\mu$ is 0.3.}
\label{mu}
\end{figure}

\subsubsection{Self-supervised Loss} We have evaluated the performance of two pretext tasks, and the results on are shown in Table \ref{ptasks}. According to the results, utilizing the absolute patch location task can approach better performance than utilizing the relative patch location task \rev{($p=0.041<0.05$)}. The reason could be that different lung regions play different roles in COVID-19 severity assessment as shown by an existing study \cite{ra78}, while the absolute patch location task may help to extract both high-level semantics and spatial information.
For the absolute patch location task, we further conduct experiments to find the best value of $\mu$, and the results are illustrated in Fig. \ref{mu}.\par
\rev{There exist some other pretext tasks in self-supervised learning, such as colorization \cite{ross2018exploiting}, denoising \cite{alex2017semisupervised}, image restoration \cite{chen2019self} and so on. However, we argue that the patch location prediction is applicable for our method, because it is patch-oriented pretext task. Considering that patches have been previously defined and cropped in the MIL setting, no more image transformation is required. Besides, in our preliminary experiments, we noted that other pretext tasks might not be applicable in severity assessment of COVID-19 due to the property of strong spatial relation and low CT contrast.}

\section{Discussion}
\rev{For future directions of our current study, we consider four following aspects to improve our work:
\begin{itemize}
    \item \textbf{Data preprocessing.} We will consider a better way of data preprocessing. In the current study, we mainly focus on designing the learning method, thus implementing the data preprocessing in a simple way. In our future work, we will incorporate automatic segmentation methods to obtain more fine-grained patches, for further performance improvement. In the meanwhile, we can also reduce a large number of irrelevant patches to further improve the effectiveness and efficiency of our method.
    \item \textbf{Self-supervised representation.} In this study, the efficacy of self-supervise learning has been evaluated. In our future work, we will further incorporate more advanced methods of self-supervised contrastive learning. In this way, we can exploit informative representation from unsupervised manner.
    \item \textbf{Longitudial information.} Longitudial information could benefit the prediction of changing trend for better severity assessment. In our future work, we will also incorporate longitudinal CT scans for severity assessment, to provide better treatment and follow-up of COVID-19 patients.
    \item \textbf{Manual delineation.} Since manual annotation is laborious, we will investigate semi-supervised learning model to further alleviate the requirement of amount of annotation.
\end{itemize} 
}\par
\rev{For possible clinical applications, we believe that our proposed method has great potential. First, by training with a small number of weak-annotated CT images, our proposed method can predict the severity of COVID-19 in a high accuracy. Second, our proposed method provides a powerful feature extractor for CT images. Specifically, the bag features are actually features of CT image, and can act as imaging attributes, which can be combined with clinical/biological attributes for other tasks (\eg, patients’ outcome prediction). Moreover, our proposed method can be extended to other problems, in which the challenges of weak annotation and insufficient data also exist, besides COVID-19 severity assessment.}
\section{Conclusion}
\label{sec5}
In this paper, we investigate a challenging clinical task of fast and accurately predicting the severity level of COVID-19. We observe two issues that may obstruct the COVID-19 severity assessment: weak annotation and insufficient data. To meet these challenges, we develop a deep attention-based MIL method combined with data augmentation and self-supervised learning. Experimental results successfully demonstrate the effectiveness of our proposed components, including 1) the MIL model for bag-level prediction, 2) the instance-level augmentation technique by generating virtual positive bags, and 3) the auxiliary self-supervised loss for extracting more discriminative features. Also, our approach shows remarkably better performance when compared with the existing methods.

\section*{Acknowledgment}
The work was supported by the National Key Research and Development Program of China (2019YFC0118300), National Natural Science Foundation of China (61673203, 81927808). 

The work was also supported by the Key Emergency Project of Pneumonia Epidemic of novel coronavirus infection (2020SK3006), Emergency Project of Prevention and Control for COVID-19 of Central South University (160260005) and Foundation from Changsha Scientific and Technical Bureau, China (kq2001001).

\balance
\bibliographystyle{ieeetr}
\bibliography{refs}

\begin{thebibliography}{10}

\bibitem{ra78}
Z.~Tang, W.~Zhao, X.~Xie, Z.~Zhong, F.~Shi, J.~Liu, and D.~Shen, ``Severity
  assessment of coronavirus disease 2019 (covid-19) using quantitative features
  from chest ct images,'' {\em arXiv preprint arXiv:2003.11988}, 2020.

\bibitem{yang2020chest}
R.~Yang, X.~Li, H.~Liu, Y.~Zhen, X.~Zhang, Q.~Xiong, Y.~Luo, C.~Gao, and
  W.~Zeng, ``Chest ct severity score: an imaging tool for assessing severe
  covid-19,'' {\em Radiology: Cardiothoracic Imaging}, vol.~2, no.~2,
  p.~e200047, 2020.

\bibitem{shan2020abnormal}
F.~Shan, Y.~Gao, J.~Wang, W.~Shi, N.~Shi, M.~Han, Z.~Xue, D.~Shen, and Y.~Shi,
  ``Abnormal lung quantification in chest ct images of covid-19 patients with
  deep learning and its application to severity prediction,'' {\em Med. Phys.},
  2020.

\bibitem{li2020ct}
K.~Li, Y.~Fang, W.~Li, C.~Pan, P.~Qin, Y.~Zhong, X.~Liu, M.~Huang, Y.~Liao, and
  S.~Li, ``Ct image visual quantitative evaluation and clinical classification
  of coronavirus disease (covid-19),'' {\em European radiology}, pp.~1--10,
  2020.

\bibitem{chao2020integrative}
H.~Chao, X.~Fang, J.~Zhang, F.~Homayounieh, C.~D. Arru, S.~R. Digumarthy,
  R.~Babaei, H.~K. Mobin, I.~Mohseni, L.~Saba, {\em et~al.}, ``Integrative
  analysis for covid-19 patient outcome prediction,'' {\em Medical Image
  Analysis}, vol.~67, p.~101844, 2020.

\bibitem{chassagnon2020ai}
G.~Chassagnon, M.~Vakalopoulou, E.~Battistella, S.~Christodoulidis, T.-N.
  Hoang-Thi, S.~Dangeard, E.~Deutsch, F.~Andre, E.~Guillo, N.~Halm, {\em
  et~al.}, ``Ai-driven quantification, staging and outcome prediction of
  covid-19 pneumonia,'' {\em Medical Image Analysis}, vol.~67, p.~101860, 2020.

\bibitem{zhou2018brief}
Z.-H. Zhou, ``A brief introduction to weakly supervised learning,'' {\em
  National Science Review}, vol.~5, no.~1, pp.~44--53, 2018.

\bibitem{rb1}
J.~Amores, ``Multiple instance classification: Review, taxonomy and comparative
  study,'' {\em Artificial intelligence}, vol.~201, pp.~81--105, 2013.

\bibitem{rb23}
P.~O. Pinheiro and R.~Collobert, ``From image-level to pixel-level labeling
  with convolutional networks,'' in {\em CVPR}, pp.~1713--1721, 2015.

\bibitem{rb30}
J.~Wu, Y.~Yu, C.~Huang, and K.~Yu, ``Deep multiple instance learning for image
  classification and auto-annotation,'' in {\em CVPR}, pp.~3460--3469, 2015.

\bibitem{rb3}
B.~Babenko, N.~Verma, P.~Doll{\'a}r, and S.~J. Belongie, ``Multiple instance
  learning with manifold bags,'' in {\em ICML}, 2011.

\bibitem{rb27}
M.~Sun, T.~X. Han, M.-C. Liu, and A.~Khodayari-Rostamabad, ``Multiple instance
  learning convolutional neural networks for object recognition,'' in {\em 2016
  23rd International Conference on Pattern Recognition (ICPR)}, pp.~3270--3275,
  IEEE, 2016.

\bibitem{cheplygina2019not}
V.~Cheplygina, M.~de~Bruijne, and J.~P. Pluim, ``Not-so-supervised: a survey of
  semi-supervised, multi-instance, and transfer learning in medical image
  analysis,'' {\em Medical image analysis}, vol.~54, pp.~280--296, 2019.

\bibitem{rbQ}
G.~Quellec, G.~Cazuguel, B.~Cochener, and M.~Lamard, ``Multiple-instance
  learning for medical image and video analysis,'' {\em IEEE reviews in
  biomedical engineering}, vol.~10, pp.~213--234, 2017.

\bibitem{rbS}
K.~Sirinukunwattana, S.~E.~A. Raza, Y.-W. Tsang, D.~R. Snead, I.~A. Cree, and
  N.~M. Rajpoot, ``Locality sensitive deep learning for detection and
  classification of nuclei in routine colon cancer histology images,'' {\em
  IEEE transactions on medical imaging}, vol.~35, no.~5, pp.~1196--1206, 2016.

\bibitem{admil}
M.~Ilse, J.~Tomczak, and M.~Welling, ``Attention-based deep multiple instance
  learning,'' in {\em International Conference on Machine Learning},
  pp.~2127--2136, 2018.

\bibitem{ad3dmil}
Z.~{Han}, B.~{Wei}, Y.~{Hong}, T.~{Li}, J.~{Cong}, X.~{Zhu}, H.~{Wei}, and
  W.~{Zhang}, ``Accurate screening of covid-19 using attention-based deep 3d
  multiple instance learning,'' {\em IEEE Transactions on Medical Imaging},
  vol.~39, no.~8, pp.~2584--2594, 2020.

\bibitem{rbK}
M.~Kandemir, M.~Haussmann, F.~Diego, K.~T. Rajamani, J.~Van Der~Laak, and F.~A.
  Hamprecht, ``Variational weakly supervised gaussian processes.,'' in {\em
  BMVC}, pp.~71.1--71.12, 2016.

\bibitem{rbH}
L.~Hou, D.~Samaras, T.~M. Kurc, Y.~Gao, J.~E. Davis, and J.~H. Saltz,
  ``Patch-based convolutional neural network for whole slide tissue image
  classification,'' in {\em CVPR}, pp.~2424--2433, 2016.

\bibitem{rc63}
L.~Taylor and G.~Nitschke, ``Improving deep learning with generic data
  augmentation,'' in {\em 2018 IEEE Symposium Series on Computational
  Intelligence (SSCI)}, pp.~1542--1547, IEEE, 2018.

\bibitem{mixup}
H.~Zhang, M.~Cisse, Y.~N. Dauphin, and D.~Lopez-Paz, ``mixup: Beyond empirical
  risk minimization,'' {\em arXiv preprint arXiv:1710.09412}, 2017.

\bibitem{rc70}
Z.~Zhong, L.~Zheng, G.~Kang, S.~Li, and Y.~Yang, ``Random erasing data
  augmentation.,'' in {\em AAAI}, pp.~13001--13008, 2020.

\bibitem{rc71}
T.~DeVries and G.~W. Taylor, ``Improved regularization of convolutional neural
  networks with cutout,'' {\em arXiv preprint arXiv:1708.04552}, 2017.

\bibitem{rc49}
M.~Frid-Adar, E.~Klang, M.~Amitai, J.~Goldberger, and H.~Greenspan, ``Synthetic
  data augmentation using gan for improved liver lesion classification,'' in
  {\em 2018 IEEE 15th international symposium on biomedical imaging (ISBI
  2018)}, pp.~289--293, IEEE, 2018.

\bibitem{rc36}
L.~Perez and J.~Wang, ``The effectiveness of data augmentation in image
  classification using deep learning,'' {\em arXiv preprint arXiv:1712.04621},
  2017.

\bibitem{rc37}
J.~Lemley, S.~Bazrafkan, and P.~Corcoran, ``Smart augmentation learning an
  optimal data augmentation strategy,'' {\em Ieee Access}, vol.~5,
  pp.~5858--5869, 2017.

\bibitem{cubuk2019autoaugment}
E.~D. Cubuk, B.~Zoph, D.~Mane, V.~Vasudevan, and Q.~V. Le, ``Autoaugment:
  Learning augmentation strategies from data,'' in {\em CVPR}, pp.~113--123,
  2019.

\bibitem{Larsson16}
G.~Larsson, M.~Maire, and G.~Shakhnarovich, ``Learning representations for
  automatic colorization,'' in {\em European conference on computer vision},
  pp.~577--593, Springer, 2016.

\bibitem{Zhang16}
R.~Zhang, P.~Isola, and A.~A. Efros, ``Colorful image colorization,'' in {\em
  European conference on computer vision}, pp.~649--666, Springer, 2016.

\bibitem{Doersch15}
C.~Doersch, A.~Gupta, and A.~A. Efros, ``Unsupervised visual representation
  learning by context prediction,'' in {\em ICCV}, pp.~1422--1430, 2015.

\bibitem{Noroozi16}
M.~Noroozi and P.~Favaro, ``Unsupervised learning of visual representations by
  solving jigsaw puzzles,'' in {\em European Conference on Computer Vision},
  pp.~69--84, Springer, 2016.

\bibitem{Gidaris18}
S.~Gidaris, P.~Singh, and N.~Komodakis, ``Unsupervised representation learning
  by predicting image rotations,'' {\em arXiv preprint arXiv:1803.07728}, 2018.

\bibitem{Pathak16}
D.~Pathak, P.~Krahenbuhl, J.~Donahue, T.~Darrell, and A.~A. Efros, ``Context
  encoders: Feature learning by inpainting,'' in {\em CVPR}, pp.~2536--2544,
  2016.

\bibitem{he2020momentum}
K.~He, H.~Fan, Y.~Wu, S.~Xie, and R.~Girshick, ``Momentum contrast for
  unsupervised visual representation learning,'' in {\em Proceedings of the
  IEEE/CVF Conference on Computer Vision and Pattern Recognition},
  pp.~9729--9738, 2020.

\bibitem{ZHOU2021101840}
Z.~Zhou, V.~Sodha, J.~Pang, M.~B. Gotway, and J.~Liang, ``Models genesis,''
  {\em Medical Image Analysis}, vol.~67, p.~101840, 2021.

\bibitem{Chen18}
T.~Chen, X.~Zhai, M.~Ritter, M.~Lucic, and N.~Houlsby, ``Self-supervised
  generative adversarial networks,'' {\em arXiv preprint arXiv:1811.11212},
  2018.

\bibitem{bf3s}
S.~Gidaris, A.~Bursuc, N.~Komodakis, P.~P{\'e}rez, and M.~Cord, ``Boosting
  few-shot visual learning with self-supervision,'' in {\em ICCV},
  pp.~8059--8068, 2019.

\bibitem{shorten2019survey}
C.~Shorten and T.~M. Khoshgoftaar, ``A survey on image data augmentation for
  deep learning,'' {\em Journal of Big Data}, vol.~6, no.~1, p.~60, 2019.

\bibitem{qin2020unsupervised}
T.~Qin, W.~Li, Y.~Shi, and Y.~Gao, ``Unsupervised few-shot learning via
  distribution shift-based augmentation,'' {\em arXiv preprint
  arXiv:2004.05805}, 2020.

\bibitem{lenet}
Y.~LeCun, L.~Bottou, Y.~Bengio, and P.~Haffner, ``Gradient-based learning
  applied to document recognition,'' {\em Proceedings of the IEEE}, vol.~86,
  no.~11, pp.~2278--2324, 1998.

\bibitem{gb}
X.~Glorot and Y.~Bengio, ``Understanding the difficulty of training deep
  feedforward neural networks,'' in {\em Proceedings of the international
  conference on artificial intelligence and statistics}, pp.~249--256, 2010.

\bibitem{adam}
D.~P. Kingma and J.~Ba, ``Adam: A method for stochastic optimization,'' {\em
  arXiv preprint arXiv:1412.6980}, 2014.

\bibitem{resnet}
K.~He, X.~Zhang, S.~Ren, and J.~Sun, ``Deep residual learning for image
  recognition,'' in {\em CVPR}, pp.~770--778, 2016.

\bibitem{densenet}
G.~Huang, Z.~Liu, L.~Van Der~Maaten, and K.~Q. Weinberger, ``Densely connected
  convolutional networks,'' in {\em CVPR}, pp.~4700--4708, 2017.

\bibitem{ross2018exploiting}
T.~Ross, D.~Zimmerer, A.~Vemuri, F.~Isensee, M.~Wiesenfarth, S.~Bodenstedt,
  F.~Both, P.~Kessler, M.~Wagner, B.~M{\"u}ller, {\em et~al.}, ``Exploiting the
  potential of unlabeled endoscopic video data with self-supervised learning,''
  {\em International journal of computer assisted radiology and surgery},
  vol.~13, no.~6, pp.~925--933, 2018.

\bibitem{alex2017semisupervised}
V.~Alex, K.~Vaidhya, S.~Thirunavukkarasu, C.~Kesavadas, and G.~Krishnamurthi,
  ``Semisupervised learning using denoising autoencoders for brain lesion
  detection and segmentation,'' {\em Journal of Medical Imaging}, vol.~4,
  no.~4, p.~041311, 2017.

\bibitem{chen2019self}
L.~Chen, P.~Bentley, K.~Mori, K.~Misawa, M.~Fujiwara, and D.~Rueckert,
  ``Self-supervised learning for medical image analysis using image context
  restoration,'' {\em Medical image analysis}, vol.~58, p.~101539, 2019.

\end{thebibliography}

\end{document}